\documentclass[pre,aps,superscriptaddress,preprint,floatfix,nofootinbib]{revtex4-1}

\usepackage[super]{nth}
\usepackage[ruled,vlined]{algorithm2e}

\usepackage{amssymb,mathrsfs,amsfonts}
\usepackage{amsmath}
\usepackage{epstopdf}
\usepackage{nicefrac}
\usepackage{listings}
\usepackage{upgreek}
\usepackage{physics}
\usepackage{multirow}
\usepackage{color}
\usepackage{natbib}
\usepackage{textcomp}
\usepackage{bibnames}
\usepackage{appendix}
\usepackage{nicefrac}
\usepackage{listings}
\usepackage{upgreek}
\usepackage{physics}
\usepackage{multirow}
\usepackage{color}
\usepackage{graphicx}
\usepackage{tabls}
\usepackage{afterpage}
\usepackage{amsthm}
\usepackage{lastpage}
\usepackage{empheq, etoolbox}
\usepackage{epstopdf}
\usepackage{changes}

\usepackage{scalerel}
\usepackage{tikz}
\usetikzlibrary{svg.path}
\usepackage{orcidlink}

\begin{document}

\title{Frequency-dependent Discrete Implicit Monte-Carlo Scheme for the Radiative Transfer Equation} 

\author
{Elad Steinberg\orcidlink{0000-0003-0053-0696}}
\email{elad.steinberg@mail.huji.ac.il}
\affiliation{Racah Institute of Physics, The Hebrew University, 9190401 Jerusalem, Israel}
\author
{Shay I. Heizler\orcidlink{0000-0002-9334-5993}}
\email{shay.heizler@mail.huji.ac.il}
\affiliation{Racah Institute of Physics, The Hebrew University, 9190401 Jerusalem, Israel}

\begin{abstract}
This work generalizes the discrete implicit Monte-Carlo (DIMC) method for modeling the radiative transfer equation from a gray treatment to an frequency-dependent one. The classic implicit Monte-Carlo (IMC) algorithm, that has been used for several decades, suffers from a well-known numerical problem, called teleportation, where the photons might propagate faster than the exact solution due to the finite size of the spatial and temporal resolution. The Semi-analog Monte-Carlo algorithm proposed the use of two kinds of particles, photons and material particles that are born when a photon is absorbed. The material particle can `propagate' only by transforming into a photon, due to black-body emission. While this algorithm produces a teleportation-free result, it is noisier results compared to IMC due to the discrete nature of the absorption-emission process. In a previous work [Steinberg and Heizler, ApJS, 258:14 (2022)], proposed a gray version of DIMC, that makes use of two kinds of particles, and therefore has teleportation-free results, but also uses the continuous absorption algorithm of IMC, yielding smoother results. This work is a direct frequency-dependent (energy-dependent) generalization of the DIMC algorithm. We find in several one and two dimensional benchmarks, that the new frequency-dependent DIMC algorithm yields teleportation-free results on one hand, and smooth results with IMC-like noise level.
\end{abstract}

\maketitle

\section{Introduction}

The radiative transfer equation (RTE) is the key equation for modeling the transport of photons that interact with the matter, calculating the specific intensity of the phase space density of photons~\cite{pomraning1973,castor2004}. Solving the RTE is crucial for understanding many astrophysical phenomena (e.g. supernova, shock breakout etc.)~\cite{sn,break}, as well as the modeling of inertial confinement fusion (ICF)~\cite{ICF} and in general high energy density physics (HEDP)~\cite{zeldovich}. 

The RTE is the Boltzmann equation for photons, where the main physical events are absorption and black-body emission through the opacity of the material, which is usually a function of the temperature and the density of the material (under the assumption that the matter itself is in local-thermodynamic equilibrium, and the electrons are distributed with the Maxwell-Boltzmann distribution with a given temperature)~\cite{pomraning1973}. In addition, scattering terms may be involved also, which can be elastic like the Thomson approximation or inelastic using the full Compton treatment~\cite{pomraning1973,castor2004}. 

The Boltzmann equation is an integro-differential equation, where in the three-dimensions the phase space contains 7 independent variables, 3 spatial, 3 velocity components (usually replaced by energy and direction) and time~\cite{pomraning1973,Duderstadt}. A full analytic solution is rare, and can be achieved in very simplified problems. For general purposes, and specifically in higher dimensions, numerical solution are used, where the two main approaches is solving numerically the deterministic Boltzmann equation for photons, or using statistical approaches, such as Monte-Carlo techniques. The most common approximations for solving the deterministic Boltzmann equation are the spherical harmonics method (also known as the $P_N$ approximation), where the intensity is decomposed into its moments~\cite{pomraning1973,Duderstadt}, and the discrete ordinates method (also known as the $S_N$ method) where the intensity is solved on some discrete selected ordinates~\cite{pomraning1973,Duderstadt}. The statistical approaches use a particle based Monte-Carlo (MC) approach, where the most common one for radiative transfer is called the implicit Monte-Carlo (IMC) method, by Fleck and Cummings~\cite{IMC,four_decades}. 

The major innovation in IMC was the implicitization of the Boltzmann equation~\cite{IMC}. Since the radiation is coupled to the matter via black-body emission, the explicit way to solve the radiation transport is to use the temperature from the beginning of the time-step. In optically-thin media, this assumption is reasonable, where the matter and the radiation may differ significantly. However, in optically-thick media, where the radiation and matter energy may be both large, but similar to each other, there is a need for very small time-steps in order to keep the solution stable. Fleck and Cummings~\cite{IMC} have offered an implicit version of radiative Monte-Carlo by finite-differencing the matter energy equation, and substituting it into the radiation Boltzmann equation. This enables the use of sufficiently large time steps, even in optically thick media, and enables the radiative transfer to be modeled using reasonable computer resources. The second novelty of IMC is the use of implicit (or continuous) absorption algorithm~\cite{IMC,milagro,four_decades}. Instead of treating the mechanism of absorption as discrete statistically events, the photon deposits its energy along its track in the spatial cells. This algorithm enables significantly smoother results than the na\"ive explicit MC. The IMC algorithm, with some modifications and improvements is being used during the last five decades~\cite{four_decades,Gentile2001,Densmore2011,teleportation,teleportation2,Gentile2010,Gentile2011,Gentile2012,Gentile2014,MCCLARREN_URBATSCH,shi_2020_1,shi_2020_3,shi_2019,gael2,tilt, Wollaeger2016, Smedley-Stevenson,Densmore}.

However, IMC suffers from a well-known numerical disadvantage, called teleportation, where the photons propagate faster than they should due to finite spatial resolution and time-steps~\cite{teleportation,teleportation2,ISMC}. In the simplest IMC implementation, when photons reach a cold opaque cell, they are absorbed in the outer part of the cell, due to the small mean free path (mfp), changing slightly the temperature of the cell. In the next time-step, the photons that are emitted due to black-body emission are sampled uniformly from the whole volume of the cell, which causes artificial propagation. Moreover, at a finite spatial resolution, this may lead to a certain `magic' time-step in which the results are correct, while deceasing the time-step further, increases the numerical error~\cite{ISMC,Steinberg1}. One elegant way to solve this problem, called semi-analog MC (SMC) and proposed by Ahrens and Larsen~\cite{SMC}, uses two kind of particles, photons and matter-particles. A photon propagates with the speed of light $c$ in the medium until it is absorbed, where it becomes a material particle (with zero-velocity). The material particle can transform into a photon due to a black-body emission process. In this way, the creation of photons can only be from the positions that they were absorbed, and the teleportation errors are avoided. Unfortunately, this scheme is explicit, and thus, enforces a small time-step when involving optically thick regions.

Recently, in a seminal work, Poëtte and Valentin offered an implicit version of SMC, ISMC, that enables the use of sufficiently large time steps~\cite{ISMC}. This new method offers a new implicitization technique that changes the RTE and the matter energy equation to be linear in both the radiation specific intensity and the matter-energy, and can be modeled via Monte-Carlo techniques using photons and material particles. The method was verified against one-dimensional gray benchmarks involving opaque media, yielding results that did not suffer from teleportation errors. Later, the ISMC method was generalized to frequency-dependent problems, yielding excellent results against many benchmarks, gray and frequency-dependent, both in 1D and 2D geometries (in both XY and RZ symmetries)~\cite{Steinberg1}. 

Another important work was published recently by Poëtte, Valentin and Bernede, that cancels the teleportation error within the frame of legacy (gray) IMC codes, called nssIMC (for no-source-sampling  IMC)~\cite{gael2}. The work presents the very few modifications that have to be done in legacy IMC codes, where the classic source sampling is canceled, and the source events are treated as `collisions events', thus teleportation is avoided. However, nssIMC converges slower than ISMC in both the spatial resolution and finite time steps (please see Fig.~3 in~\cite{gael2}). In addition, nssIMC produces noisier results in a given resolution comparing to ISMC (Fig.~4 in~\cite{gael2}) and is less robust. Therefore, where the ISMC modifications are allowable, the authors recommend to opt for ISMC rather than nssIMC.

However, both SMC and ISMC yield noisier results compared to IMC, due to the discrete treatment of the absorption and black-body emission processes. In addition, since the total number of particles, both photons and material particles is constant, problems that have a big difference between the heat-capacities of the radiation and the material requires a large statistic to converge. Recently a new algorithm was offered, called discrete implicit Monte-Carlo (DIMC)~\cite{Steinberg2} that uses the idea of two kinds of particles on one hand, and thus yields teleportation-free results, and uses the algorithm, of continuous (implicit) absorption on the other hand, yielding smooth results as IMC. This is attainable due to an algorithm that avoids the population of particles from exploding, and retaining the correct angular and spatial distributions. DIMC was derived using the IMC linearization, but this wasn't a mandatory choice, it could have been derived via the ISMC linearization and implicitization as well. This algorithm was tested against several {\em{gray}} benchmarks, both in 1D (with or without hydrodynamics) and 2D. In this work we expand the DIMC method to include frequency-dependent problems. The new frequency-dependent DIMC method will be examined against all the the frequency-dependent benchmarks that were tested using ISMC in~\cite{Steinberg1}. We repeat the frequency-dependent benchmarks from~\cite{Steinberg1}, adding to each figure the comparable DIMC result. 

This work is structured as follows: In Sec.~\ref{DIMC_MG} we present the frequency-dependent version of DIMC. In Sec.~\ref{1d} we present several 1D frequency-dependent benchmarks, while in Sec.~\ref{2d} we present a 2D frequency-dependent benchmark.  We finish with a short conclusions section in Sec.~\ref{discussion}.

\section{Frequency-dependent DIMC}
\label{DIMC_MG}

The basic derivation is the classic derivation of IMC implicitization and linearization.
For the case of elastic (Thomson) scattering, the RTE and the coupled energy-balance equation for the material energy are~\cite{pomraning1973}:
\begin{subequations}
\label{basic}
\begin{align}
\label{RTE1}
    \frac{1}{c}\frac{\partial I(\boldsymbol{r},\boldsymbol{\Omega},\nu,t)}{\partial t}+\boldsymbol{\Omega\cdot\nabla} I(\boldsymbol{r},\boldsymbol{\Omega},\nu,t)&=
    \begin{aligned}[t]
    &-(\sigma_{a}(\nu,T)+\sigma_{s}(\nu,T))I(\boldsymbol{r},\boldsymbol{\Omega},\nu,t)+\sigma_{a}(\nu,T)B(\nu,T)\\
    &+\sigma_{s}(\nu,T)\int_{4\pi}\frac{I_\nu(\boldsymbol{r},\boldsymbol{\Omega}',t)}{4\pi}\boldsymbol{d\Omega}'\\
    \end{aligned}\\
    \frac{\partial e(T)}{\partial t}&=c \int_0^{\infty}\left[\sigma_{a}(\nu',T)\left(\int_{4\pi}I(\boldsymbol{r},\boldsymbol{\Omega}',\nu',t)\boldsymbol{d\Omega}'-4\pi B(\nu',T)\right)d\nu' \right],
    \label{matter1}
    \end{align}
\end{subequations}
where $I(\boldsymbol{r},\boldsymbol{\Omega},\nu,t)$ is the radiation specific intensity for a unit space $\boldsymbol{r}$, unit direction $\boldsymbol{\Omega}$ and unit frequency $\nu$ at time $t$, $\sigma_a(\nu,T)$ is the absorption cross-section (opacity) which is a function of the material temperature $T$ and the frequency, $\sigma_s(\nu,T)$ is the scattering cross-section, $e(T)$ is the thermal energy per unit volume of the medium, $c$ is the speed of light and $B(\nu,T)=\frac{2h\nu^3}{c^2}\left(\exp{(h\nu/k_BT)}-1\right)^{-1}$ is the Planck function ($h$ is the Planck constant and $k_B$ is the Boltzmann constant).

Note that the frequency integration over $B(\nu,T)$ is called the {\em{frequency integrated}} Planck function which equals to $B(T)=aT^4/4\pi$ and $a$ is the radiation constant. We define a convenient function which is the ratio of the Planck function to the frequency integrated one $b(\nu,T)\equiv B(\nu,T)/B(T)=4\pi B(\nu,T)/aT^4$.

The basic idea behind IMC, is to define the Fleck parameter,
\begin{equation}
    f=\frac{1}{1+\beta \sigma_{a,P}(T) c  \Delta t}
\end{equation}
where $\beta\equiv\partial aT^4/\partial e$ is the ratio between the radiation and material heat capacities~\cite{IMC} (which is taken at the beginning of the time-step), $\Delta t$ is the discretized time-step and $\sigma_{a,p}(T)$ is the Planck opacity $\sigma_{a,p}(T)\equiv\int_0^{\infty}\sigma_{a}(\nu,T)b_\nu d\nu$ (also taken at the beginning of the time-step). In a matter of fact, taking the heat-capacity and the opacities at the beginning of the time-step prevents the classic implementation of IMC to be a fully implicit algorithm~\cite{Gentile2011,Gentile2014}. Finite-differencing Eq.~\ref{matter1} and using the definition of $\beta$ yields (where the index $n$ denotes the previous time step and $n+1$ the current time step):
\begin{equation}
    B^{n+1}(T)=B^{n}(T)f+(1-f)\int_0^{\infty}\int_{4\pi}\frac{\sigma_{a,\nu'}}{\sigma_{a,p}(T)}\frac{I(\boldsymbol{r},\boldsymbol{\Omega}',\nu',t)}{4\pi}\boldsymbol{d\Omega}'d\nu'
\label{fleck_energy}
\end{equation}
i.e., $B^n(T)=a{(T^n)}^4/4\pi$ and $b^n(\nu,T)\equiv B^n(\nu,T)/B^n(T)=4\pi B^n(\nu,T)/a{(T^n)}^4$.

Substituting Eq.~\ref{fleck_energy} in Eqs.~\ref{basic}, yields the final frequency dependent IMC equations:
\begin{subequations}
\label{dimc_freq}
\begin{align}
    &\frac{1}{c}\frac{\partial I(\boldsymbol{r},\boldsymbol{\Omega},\nu,t)}{\partial t}+\boldsymbol{\Omega\cdot\nabla} I(\boldsymbol{r},\boldsymbol{\Omega},\nu,t)=
    -(\sigma_{a}(\nu,T)+\sigma_{s}(\nu,T))I(\boldsymbol{r},\boldsymbol{\Omega},\nu,t)\\
    &+f\sigma_{a}(\nu,T)b^n(\nu,T)B^n(T)+\sigma_{s}(\nu,T)\int_{4\pi}\frac{I_\nu(\boldsymbol{r},\boldsymbol{\Omega}',\nu',t)}{4\pi}\boldsymbol{d\Omega}'+\nonumber \\
    &(1-f)\frac{\sigma_{a}(\nu,T)b^n(\nu,T)}{\sigma_{a,p}(T)}\int_0^{\infty}\int_{4\pi}\frac{I(\boldsymbol{r},\boldsymbol{\Omega}',\nu',t)\sigma_{a}(\nu',T)}{4\pi}\boldsymbol{d\Omega}'d\nu'\nonumber
    \end{align}
    \begin{equation}
    \frac{\partial e(T)}{\partial t}=c f\left(\int_0^{\infty}\int_{4\pi}\sigma_{a}(\nu',T)I(\boldsymbol{r},\boldsymbol{\Omega}',\nu',t)\boldsymbol{d\Omega}'d\nu'-\sigma_{a,p}(T)aT^{4} \right).
    \label{dimc_matter}
    \end{equation}
\end{subequations}
These are the basic frequency-dependent IMC equations, where in addition to the physical scattering term (which is proportional to the frequency-integrated specific intensity), there is an additional, ``effective" scattering term which is proportional to $1-f$. This effective scattering term, replaces the explicit process of absorption and black-body emission, which enables the use of large time-steps, with $f<1$.

The implicitization and linearization of DIMC is the same as classic IMC, i.e., linear only in $I$ (as opposed to ISMC that is linear in both $I$ and $e$~\cite{ISMC,Steinberg1}), and the matter-energy is treated via Eq.~\ref{dimc_matter}. The difference from classic IMC is that the material energy is also represented by material particles, where the sum of the energy of the material particles inside a numerical cell is exactly the total energy of the matter in the cell. Photons are allowed to be created only in positions of material particles, and are absorbed continuously, generating new material particles. To prevent population explosion, an algorithm of merging material particles weighted by their energy was derived, and limits at the end of the time step the number of material particles to be approximately 10-30 in each cell. For a more complete description of the DIMC algorithms, the algorithm of photon creation, the single photon propagation and the algorithm of material particle merging, see~\cite{Steinberg2}. 
The only difference in the frequency-dependent algorithm is that when a new radiation photon is emitted or when there is an effective scattering event, the photon's new frequency is sampled in the same manner as in the emission case in frequency-dependent IMC and ISMC. The propagation is determined via the frequency-dependency of the opacity. We note that as in classic IMC, the Monte-Carlo particles that represent the photons are {\em{not}} photons per se, but rather `packets of energy', which represent large number of photons (calculating real number of photons is unrealistic). As a consequence, there is not a direct relation between the frequency of the energy-packet and its energy, which may represent many low (or high)-energy photons.

The main sub-steps of each time advancement in the DIMC scheme are as follows:
\begin{itemize}
    \item The temperature of a cell is calculated from $e$ (which is equal to the sum of energies from all the material particles in a given cell).
    \item The values for $f$ and $\beta$ are calculated using the values at the beginning of the time step.
    \item A given number of new photons $N_{\mathrm{photon}}$~(packets of energy) is created (a parameter by the user), where the total energy of the photons are determined by the black-body emission $V\Delta tf\int\sigma_{a,p}(T)cB(\nu,T)d\nu$ ($V$ is the volume of the cell), as classic IMC. New positions for the photons are sampled randomly from the positions of the material particles in the cell, weighted by their energy. Each packet of energy has the same energy, where its energy is then subtracted from the energy of the cell as well as from the sampled material particle. The frequency of the photon is sampled as in IMC, by randomly sampling the distribution $\int\sigma_a(\nu,T)b(\nu,T)d\nu/\sigma_{a,P}(T)$. Since the emission depends explicitly in the opacity $\sigma_a$, we build a (numerical) cumulative spectral distribution of the integral above, sampling uniformly from the opacity-Planck-weighted integrated distribution (for more details, please see section IVA in~\cite{IMC}). For the detailed algorithm of photon creation please see algorithm 1 in~\cite{Steinberg2} which appears in the appendix.
    \item If there are external radiation sources/boundary conditions, new radiation photons are created with their corresponding frequency distribution.
    \item Radiation photons are transported with a velocity $c$ where the distances for a collision $d_{\mathrm{collision}}$, moving to a neighboring cell $d_{\mathrm{mesh}}$ or free flight $d_{\mathrm{time}}$ are calculated. The minimal distance is then taken and the reduction in the energy of the photon (a packet of energy, as explained previously) is then calculated continuously by $\Delta E=E_{\mathrm{photon}}(1-\exp(-f\sigma_a(\nu)d))$. This energy difference is then saved for when a new material particle is created.
    \item If there is a collision, the photon is scattered and with a probability of $(1-f)\sigma_{a}(\nu)/((1-f)\sigma_{a}(\nu)+\sigma_{s}(\nu))$ a new frequency is sampled from the same distribution as before (i.e., only for the effective scattering).
    \item
    Once the photon exits the cell or reaches the end of the time step, the position of a new material particle is determined by averaging the photon energy loss track (see algorithm 2 in~\cite{Steinberg2} which appears in the appendix). A new material particle is created at that location and its energy is equal to the sum of the energies that were deposited by the photon. The position of a new material particle when a photon crosses the boundary between cells, is not necessarily right at the boundary, but rather is determined by the track length of the photon in the old cell, due to the photon's deposition of energy inside the old cell.
    \item
    Since each photon creates at least material particle in each time step, material particles are merged until we reach some desired number $N_{\text{material particles in cell}}$. The merging process retains the ``center of mass" (where here mass is the energy) of the material particles energy (see algorithm 3 in~\cite{Steinberg2} which appears in the appendix). We note again that the main difference between DIMC and classic IMC is in the source sampling; the total energy of the cell is a global quantity. Therefore, DIMC should be consistent at least as classic IMC, and is better at avoiding teleportation errors.
\end{itemize}

We note that we haven't presented a theoretical proof that the DIMC algorithms are unbiased. However, from the numerical examples that are presented above, we see no evidence for bias. Our main concern of a source of bias, was the material particles merging algorithm. We performed extensive numerical checks with different methods of merging the material particles. Our proposed algorithm had the least bias among all of the tested methods, that included in addition to our method also Russian Roulette and a comb sort algorithms.

\section{1D Tests}
\label{1d}
In~\cite{Steinberg1} we have collected the well-known frequency-dependent radiative transfer benchmarks that appear in the literature and tested the ISMC algorithm with them. In this section we use these benchmarks to test the new DIMC scheme in frequency-dependent scenarios in one-dimension, which includes the frequency-dependent benchmarks that was offered lately by Olson~\cite{olson2020} and by Densmore et al.~\cite{Densmore,teleportation2}, in several optical depth.

\subsection{Olson 2020 1D}

The frequency-dependent Olson 1D benchmark~\cite{olson2020} is a frequency-dependent extension setup to the {\em{optically thin}} source gray test that was introduced previously in~\cite{olson2019}. The source term is a black body with a temperature of $0.5$ keV whose spatial extent is given by $Q(x)=B(0.5\text{ keV})\exp{(-693\cdot x^3)}$, and is turned on at time $t=0$ and turned off at time $ct=2$. The initial temperature in the domain $0\leqslant x \leqslant 4.8$ (cm) is set to be $T(t=0)=0.01T_\text{ keV}$ ($T_\text{ keV}$ represents the temperature in the units of keV) and we run the simulation until a time $ct=4$ using reflective boundary conditions at both ends.

The material is composed of carbon-hydrogen foam, whose frequency-dependent opacity (in units of $\text{cm}^2\text{g}^{-1}$) is given by:
\begin{equation}
    \kappa_{a}(\nu,T)=\begin{cases}
    \text{min}(10^7,10^9(T/T_\text{ keV})^2)& h\nu<0.008\text{ keV}\\
    \frac{3\cdot 10^6 \left(0.008\text{ keV}/h\nu\right)^2}{(1+200\cdot(T/T_\text{ keV})^{1.5})}& 0.008\text{ keV}<h\nu<0.3\text{ keV}\\
    \frac{3\cdot 10^6 \left(0.008\text{ keV}/h\nu\right)^2\sqrt{0.3\text{ keV}/h\nu}}{(1+200\cdot(T/T_\text{ keV})^{1.5})}+\frac{4\cdot 10^4\left(0.3\text{ keV}/h\nu\right)^{2.5}}{1+8000(T/T_\text{ keV})^2}&
    h\nu >0.3\text{ keV}.
    \end{cases}
\end{equation}
The macroscopic absorption cross-section is given by $\sigma_{a}(\nu,T)=\rho \kappa_{a}(\nu,T)$, where $\rho=0.001$ $\text{g}/\text{cm}^3$. The heat capacity is given by:
\begin{subequations}
\label{heat_cap}
\begin{eqnarray}
    \rho C_V&=&aT_\text{ keV}^3H\left(1+\alpha+\left(T+\chi\right)\frac{\partial\alpha}{\partial T}\right)\\
    \alpha&=&\frac{1}{2}e^{-\chi/T}\left(\sqrt{1+4e^{\chi/T}}-1\right)\\
    \frac{\partial\alpha}{\partial T}&=&\frac{\chi}{T^2}\left(\alpha-1/\sqrt{1+4e^{\chi/T}}\right)
\end{eqnarray}
\end{subequations}
where $\chi =0.1T_{\text{keV}}$ and $H=0.1$.
For all of the schemes we use a spatial resolution of 128 equally spaced cells and set a constant time step of $\Delta t = 10^{-13}$. We have tried to achieve comparable noise level in the material temperature field between IMC, ISMC and DIMC. In the IMC and DIMC cases, we create $500$ new photons each time step and limit the total number to be $4\cdot 10^3$, while in the ISMC case we create $5\cdot 10^3$ new photons each time step and limit the total number of particles to be $10^5$. For the reference solution, we use the one calculated using a high order $P_N$ scheme presented in~\cite{olson2020}.

\begin{figure}
(a)\includegraphics*[width=7.5cm]{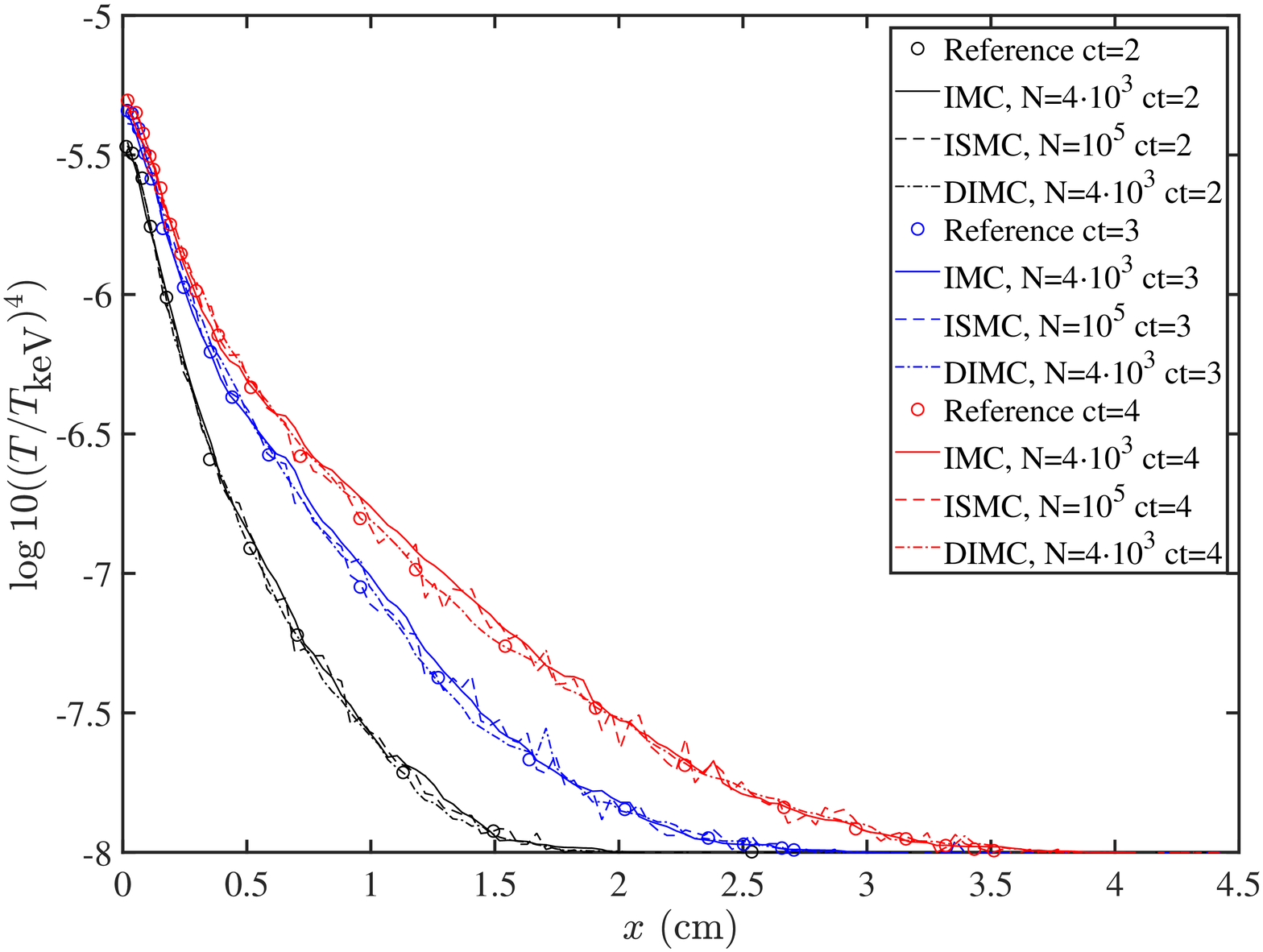}
(b)\includegraphics*[width=7.5cm]{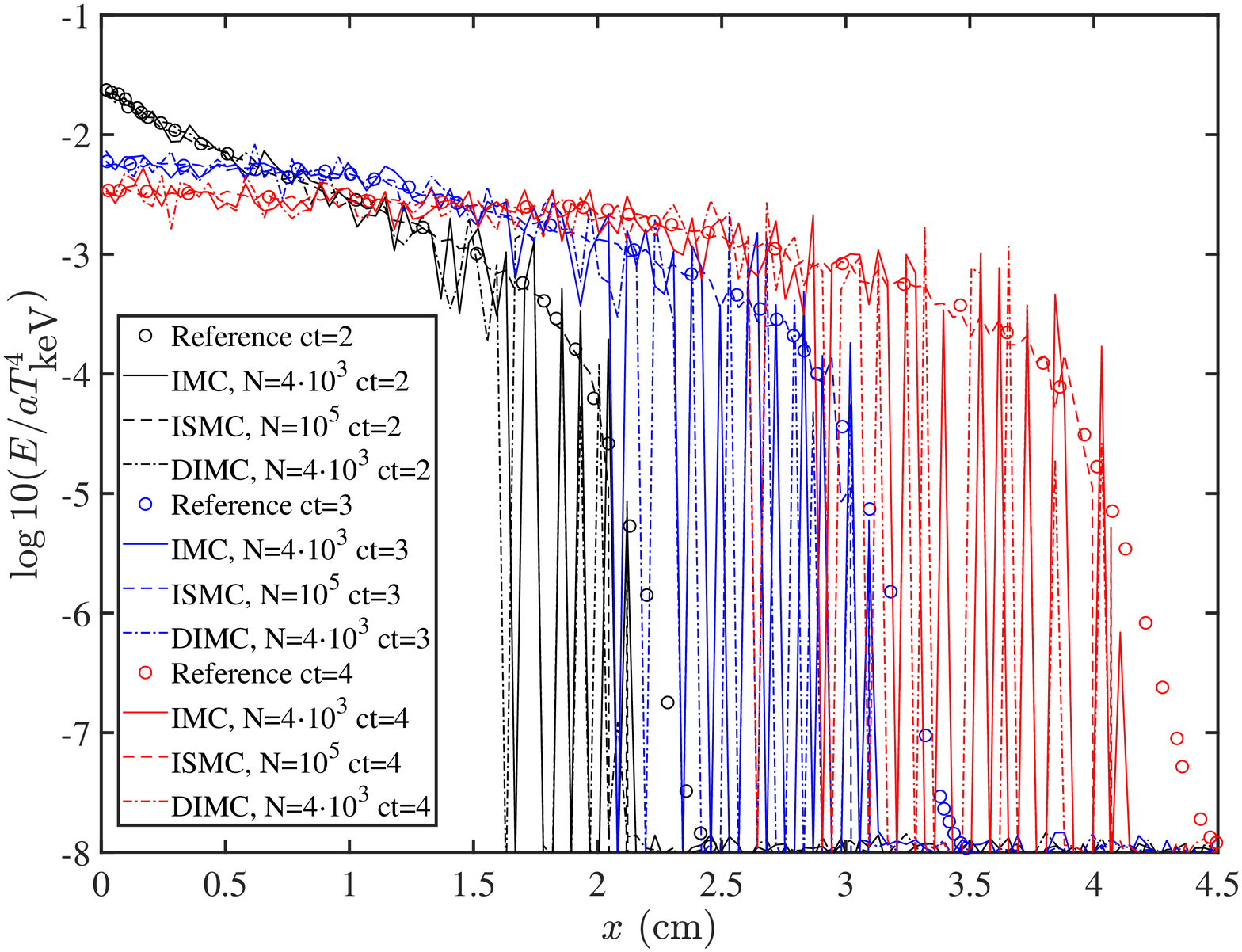}
\caption{Low statistic IMC and DIMC runs vs high statistic ISMC run for (a) The material temperature at different times for the 1D Olson~\cite{olson2020} test problem. (b) The radiation energy density at different times for the 1D Olson~\cite{olson2020} test problem.}
\label{fig:olson2020}
\end{figure}

Fig.~\ref{fig:olson2020}(a) shows the material temperature for different times compared with the reference solution, and the radiation energy density is shown in Fig.~\ref{fig:olson2020}(b).
The frequency-dependent DIMC method yields the correct result and agrees well with both the IMC and ISMC solutions and the reference solution as well. As in the ISMC results, there is a difference between the discrete results of the different MC schemes and the smooth curves of the $P_N$, near the tails of the distributions, where the radiation intensity drops by several orders of magnitudes. 
Since all three methods are converge to the same solution with an infinite number of particles, we choose for the comparison of the efficiency between the different schemes the number of particles required to achieve a {\em{similar noise level in the matter energy}} (since in most cases this quantity is more informative than the radiation energy). Thus, the noise level in the material temperature is almost comparable in all three methods while in the radiation field the noise level in ISMC is lower than both IMC ad DIMC, as expected. However, the total number of particles of the ISMC run is higher by two magnitudes than the other two methods. That is because both IMC and DIMC need relatively low statistics to yield smooth material profiles, and in ISMC the noise levels of radiation and matter are similar. This means that for a given statistic, ISMC is much noisier than IMC or DIMC. For achieving similar level of noise using ISMC needs an overall run time that was a factor of 9.5 longer than the IMC run.

\subsection{Densmore et al. 2012}
Densmore et al.~\cite{Densmore} (and also Cleveland et al.~\cite{teleportation2}) have presented one-dimensional frequency-dependent benchmarks with varying optical depth: optically thin material, medium-level (intermediate) opacity and optically thick material. In addition, they introduced a combination of two different materials or zones, i.e. the heat wave that initially propagates in optically thin material, `crashes' into an optically-thicker ``wall". The optically-thick media problems emphasize the difference between IMC, which exhibits a teleportation error, while ISMC and DIMC should be spatially converged at a lower spatial resolution with no teleportation error, even in frequency-dependent problems.

First, in the single opacity Densmore et al. benchmark, the opacity has this form:
\begin{equation}
    \sigma(x, \nu, T) = \frac{\sigma_0(x)}{\left(h\nu\right)^3\sqrt{k_BT}}
\end{equation}
where $\sigma_0(x)=10\text{keV}^{7/2}$ for the optically-thin medium, $\sigma_0(x)=100\text{keV}^{7/2}$ for the optically-intermediate medium and $\sigma_0(x)=1000\text{keV}^{7/2}$ for the optically-thick medium.
The heat capacity is set to be $C_V=10^{15}\text{erg}/T_\text{ keV}/\text{cm}^3$. The initial temperature in the domain $0\leqslant x \leqslant 5$ (cm) is 1 eV, the left boundary is a black body bath source with a temperature of 1keV and the right boundary is a reflecting wall. The spatial resolution is taken to be small, 64 evenly spaced cells (for presenting the teleportation in opaque problems in IMC). The test is run to a time of $t=1$ ns with a constant time step of $\Delta t = 0.01$ ns. For all of the runs we create $10^5$ photons per time step and limit the total number of photons to be $10^6$. The reference solution that we compare to is taken from Densmore et al.~\cite{Densmore}. 

\begin{figure}
\centering{
(a)\includegraphics*[width=6.8cm]{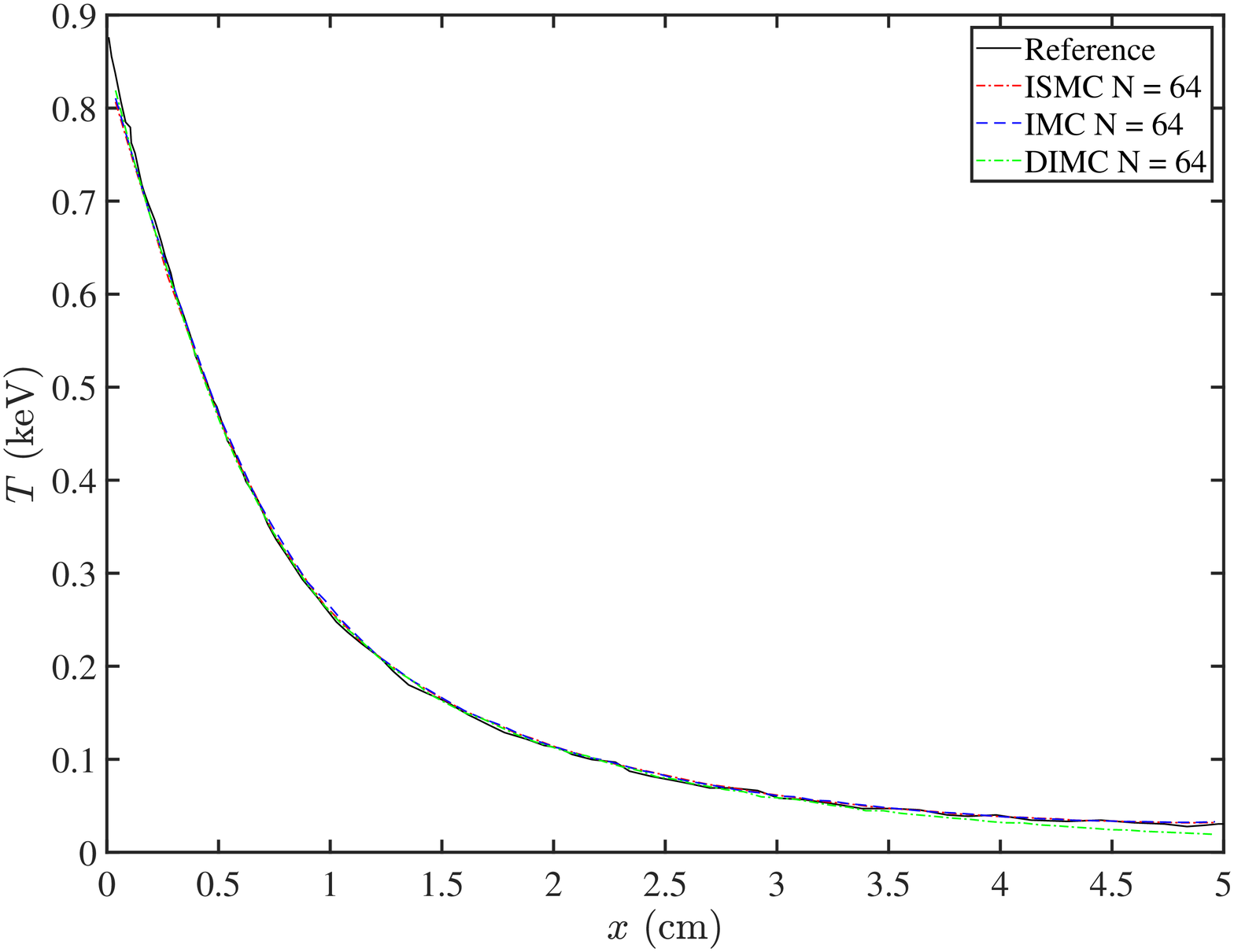}
(b)\includegraphics*[width=6.8cm]{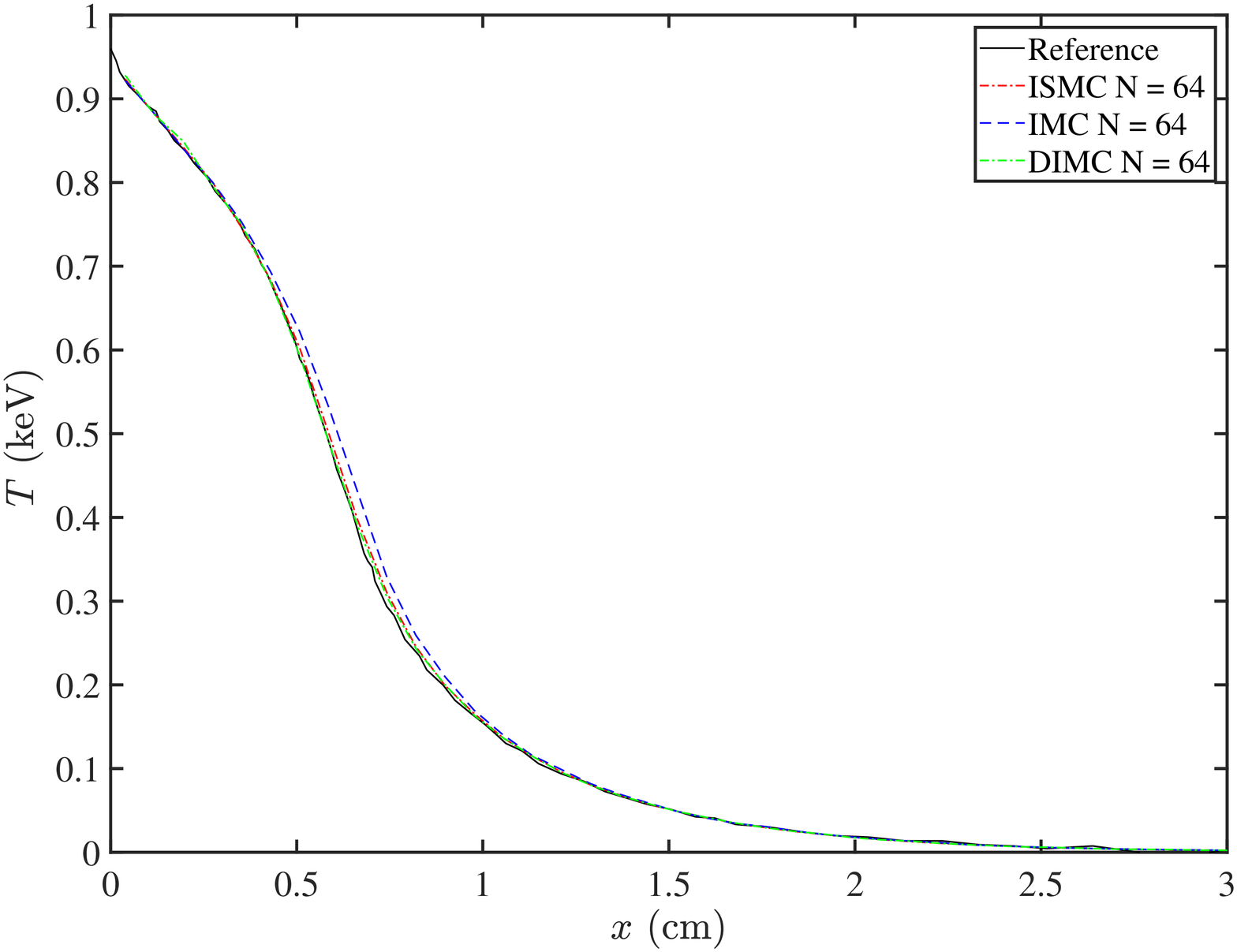}
(c)\includegraphics*[width=6.8cm]{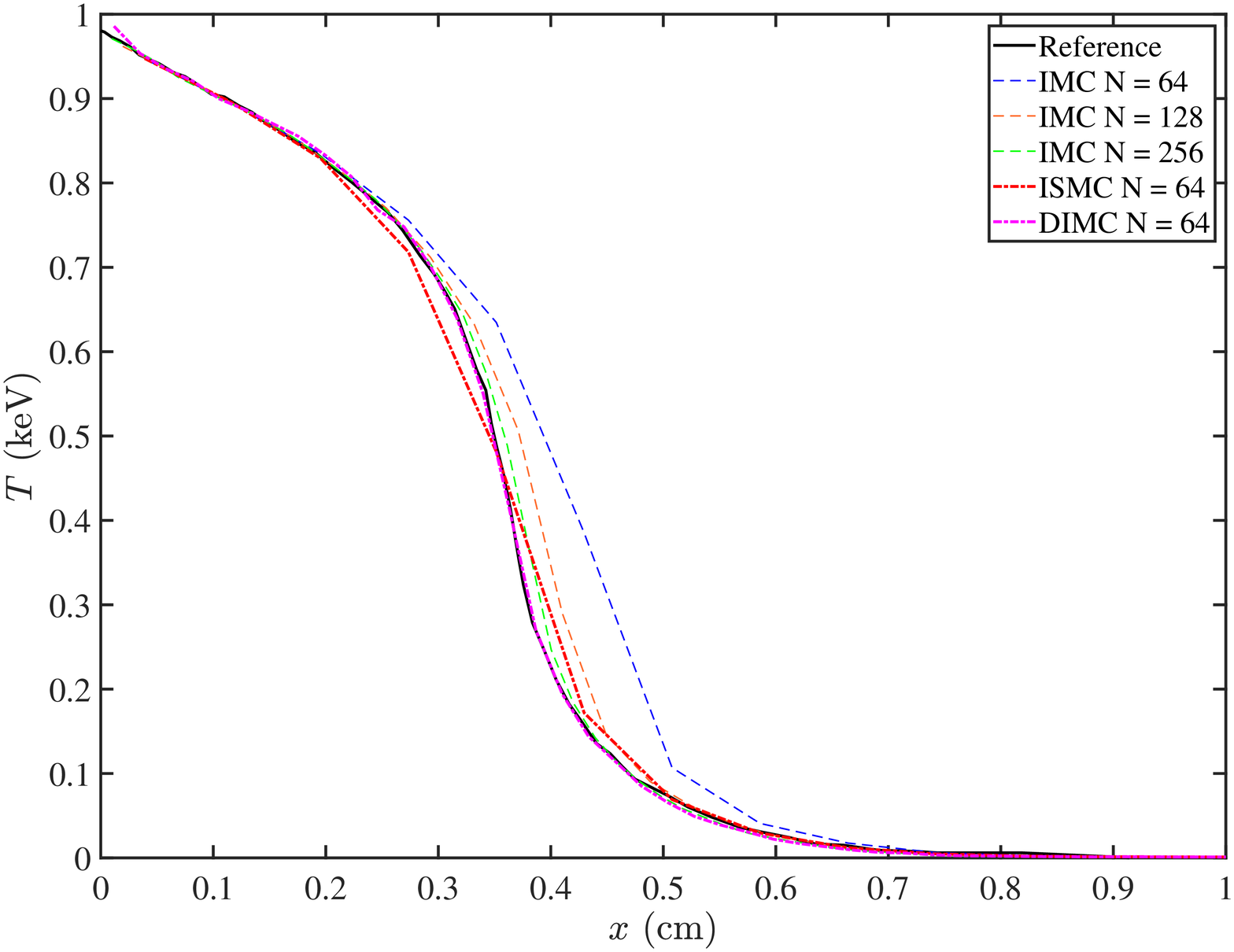}
\caption{The material temperature at time $t=1$ ns for the first three Densmore et al.~\cite{Densmore} benchmarks. (a) $\sigma_0(x)=10\;\text{keV}^{7/2}$/cm, (b) $\sigma_0(x)=100\;\text{keV}^{7/2}$/cm and (c) $\sigma_0(x)=1000\;\text{keV}^{7/2}$/cm.}
\label{fig:densmore}
}
\end{figure}

The material temperature profiles at $t=1$ ns for all three methods along with the reference solution are shown in Fig.~\ref{fig:densmore}.
We can see that there is a good agreement between the three methods and the reference solution for the first two tests (optically thin and intermediate-opacity problems), while in the third test (optically thick), the IMC requires a spatial resolution of 256 cells in order to avoid teleportation errors, as opposed to the ISMC or DIMC that give a good result even with 64 cells. As a matter of fact, the DIMC yields even better results comparing to reference solution than ISMC. Again, all three methods converge to the same solution with an infinite number of particles and spatial resolution. Therefore, in this problem we choose for the comparison of the efficiency between the different schemes, the required spatial resolution for convergence. In this metric, IMC needs four times higher resolution for achieving the same accuracy as ISMC and DIMC.

\begin{figure}
\centering
(a)\includegraphics*[width=6.8cm]{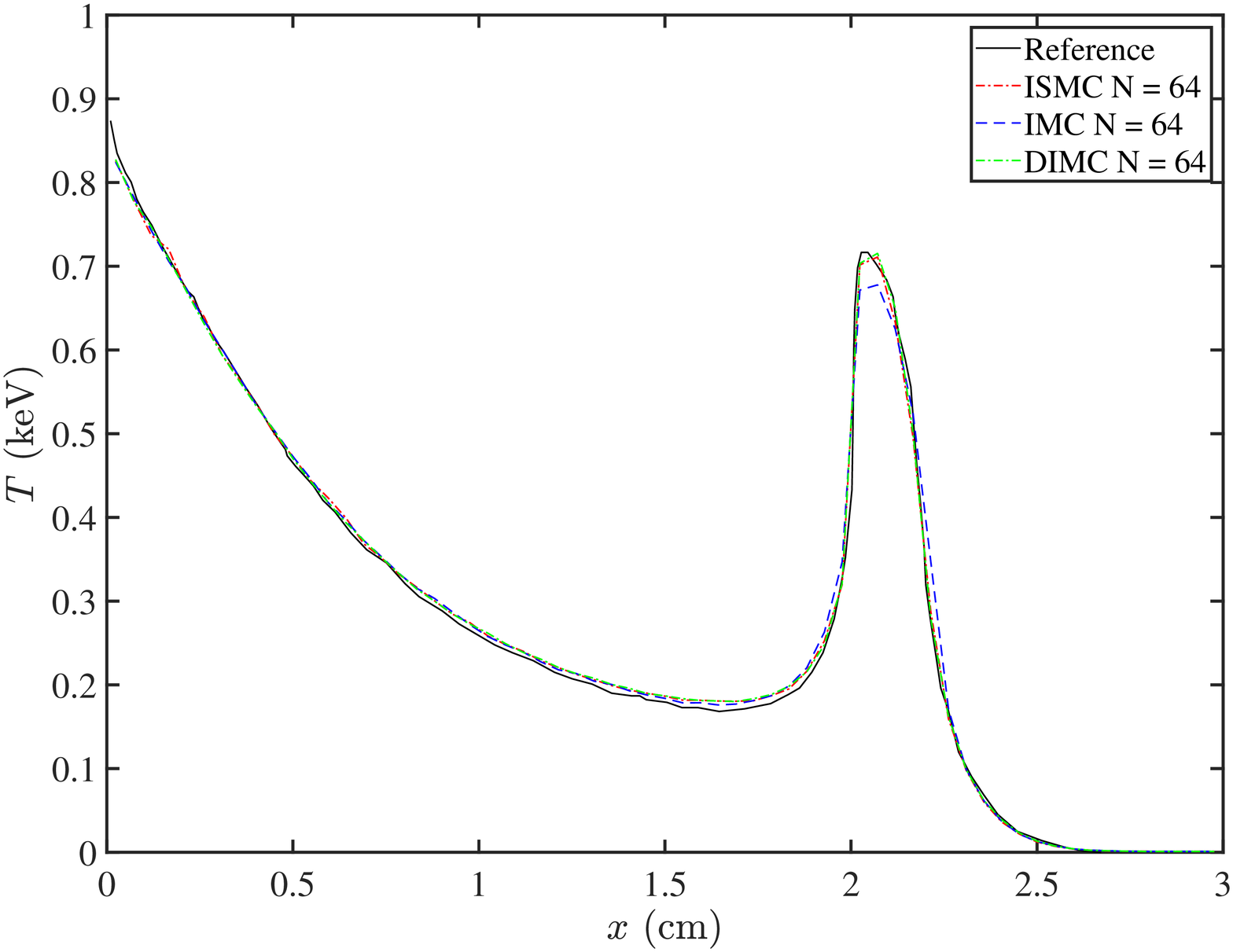}
(b)\includegraphics*[width=6.8cm]{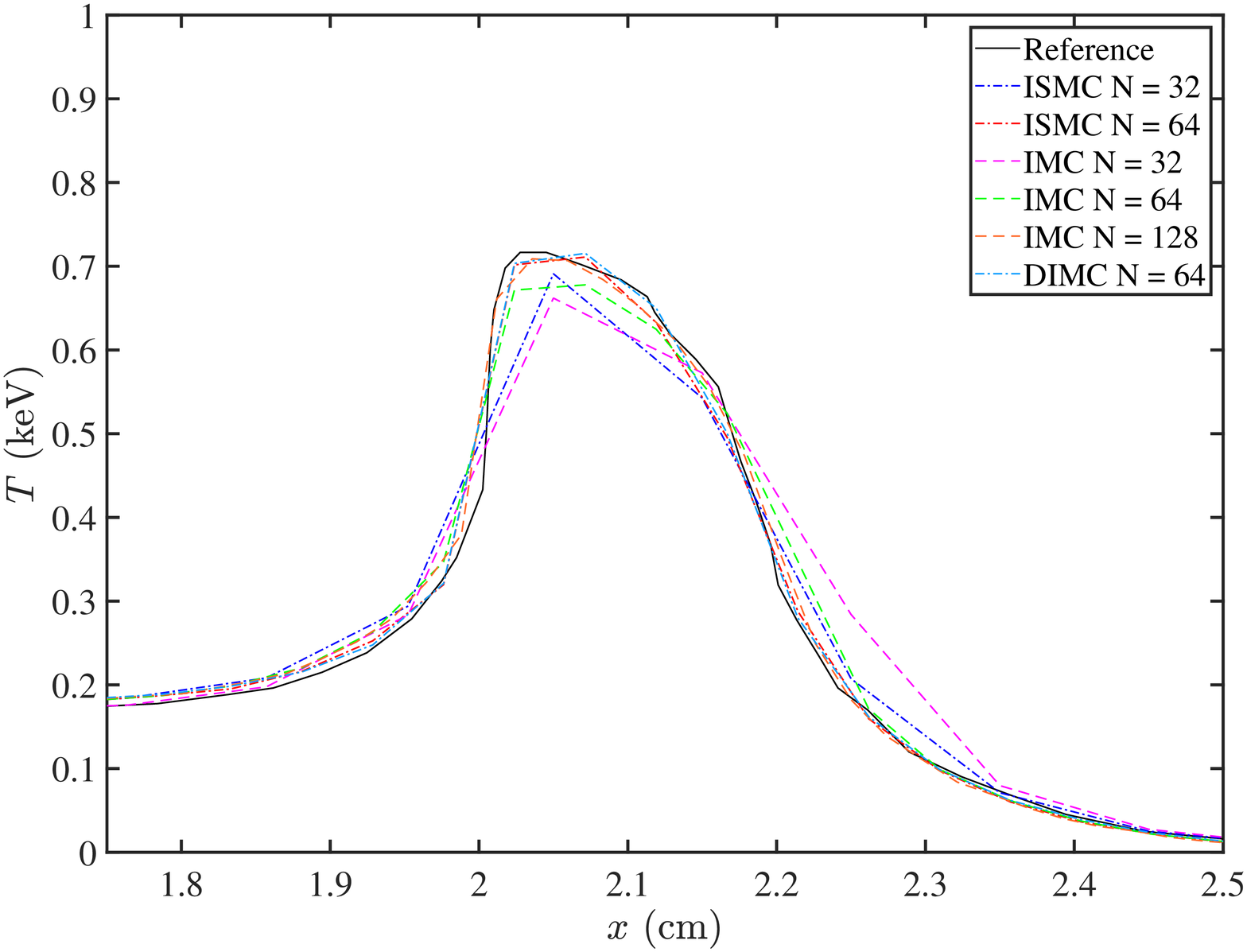}
\caption{The material temperature at time $t=1$ ns for tow-media Densmore et al.~\cite{Densmore} benchmark. (a) Zoom out showing the entire domain, (b) zoom in showing the interface between the optically thin and optically thick materials located at $x=2$ cm.}
\label{fig:densmore_interface}
\end{figure}

Next, Densmore et al. offered a two-media benchmark, testing the handling of sharp transition from an optically thin to an optically thick regime. The domain (with equally spaced cells throughout) is set to $0\leqslant x \leqslant 3$ (cm), and the opacity is
\begin{equation}
    \sigma_0(x)=\begin{cases}
    10\;\text{keV}^{7/2}/\text{cm}& x<2\;\text{cm},\\
    1000\;\text{keV}^{7/2}/\text{cm}& x\ge 2\;\text{cm}.
    \end{cases}
\end{equation}
In Fig.~\ref{fig:densmore_interface} we show the material temperature for the two-media benchmark (Fig.~\ref{fig:densmore_interface}(b) is zoomed on the interface zone). As in the optically thick single-medium benchmark, we see that when optically thick material is present, the new DIMC algorithm (as well as the ISMC algorithm) requires less spatial resolution than IMC in order to achieve convergence, due to teleportation toward the opaque material in low spatial resolution in the IMC simulation. One again, DIMC yields even better accuracy with the reference solution than ISMC.

\section{Olson 2020 2D test}
\label{2d}
\begin{figure} 
\centering
\includegraphics*[width=7.5cm]{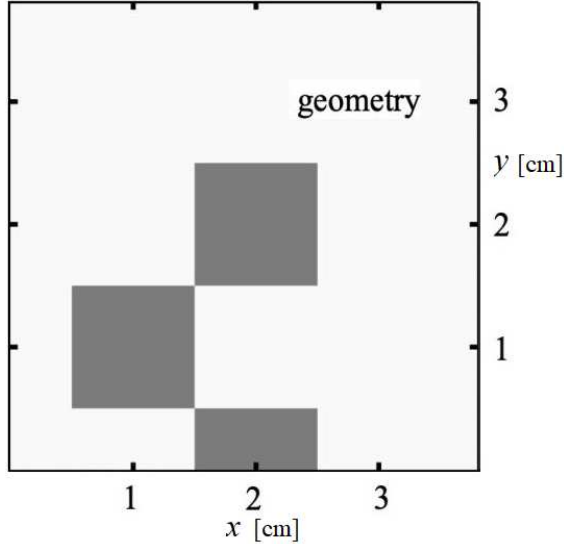}
\caption{The geometrical setup to the 2D frequency-dependent benchmark of Olson~\cite{olson2020}. The gray squares represent opaque regions.}
\label{olson20202D_setup}
\end{figure}

In this section we present a frequency-dependent implementation of the new DIMC scheme in a full two-dimensional problem, using the most complex 2D problem presented in Olson~\cite{olson2020}. In this problem, a complex-geometry 3.8 cm square on each side, composed of intermediate-opaque (thus, no teleportation errors should appear in the IMC simulations) aluminum blocks surrounded by the same foam from the previous 1D Olson benchmark problem. The exact setup is shown in Fig.~\ref{olson20202D_setup}. The aluminum blocks have the same density as the foam and their heat capacity has the same functional form as the foam (Eq.~\ref{heat_cap}), but with $H=0.5$ and $\chi=0.3T_\text{ keV}$. The opacity for the aluminum blocks is given by:
\begin{equation}
    \kappa_{a}(\nu,T)=\begin{cases}
    \text{min}(10^7,10^8T/T_\text{ keV})& h\nu<0.01\text{ keV}\\
    \frac{10^7 \left(0.01\text{ keV}/h\nu\right)^2}{(1+20\cdot(T/T_\text{ keV})^{1.5})}& 0.01\text{ keV}<h\nu<0.1\text{ keV}\\
     \frac{10^7 \left(0.01\text{ keV}/h\nu\right)^2}{(1+20\cdot(T/T_\text{ keV})^{1.5})} + \frac{10^6\left(0.1\text{ keV}/h\nu\right)^2}{1+200\cdot(T/T_\text{ keV})^2}& 0.1\text{ keV}<h\nu<1.5\text{ keV}\\
    \frac{10^7 \left(0.01\text{ keV}/h\nu\right)^2\sqrt{1.5\text{ keV}/h\nu}}{(1+20\cdot(T/T_\text{ keV})^{1.5})} + \frac{10^5\left(1.5\text{ keV}/h\nu\right)^{2.5}}{1+1000\cdot(T/T_\text{ keV})^2}&
    h\nu >1.5\text{ keV}.
    \end{cases}
\end{equation}

The computational region is divided into 380 equally spaced cells in each axis with reflecting boundaries. A black-body source, $Q(r)=B(0.5\text{ keV})\exp{(-18.7\cdot r^3)}$, is turned on at time $t=0$, and remains on for the entire duration of the simulation. A constant time step of $\Delta t = 10^{-13}$ s is used, $2\cdot 10^6$ new particles are created each time step, and we limit the total number of particles to be $2.5\cdot 10^7$ for all of the schemes.

\begin{figure} 
\centering
(a)\includegraphics*[width=6.8cm]{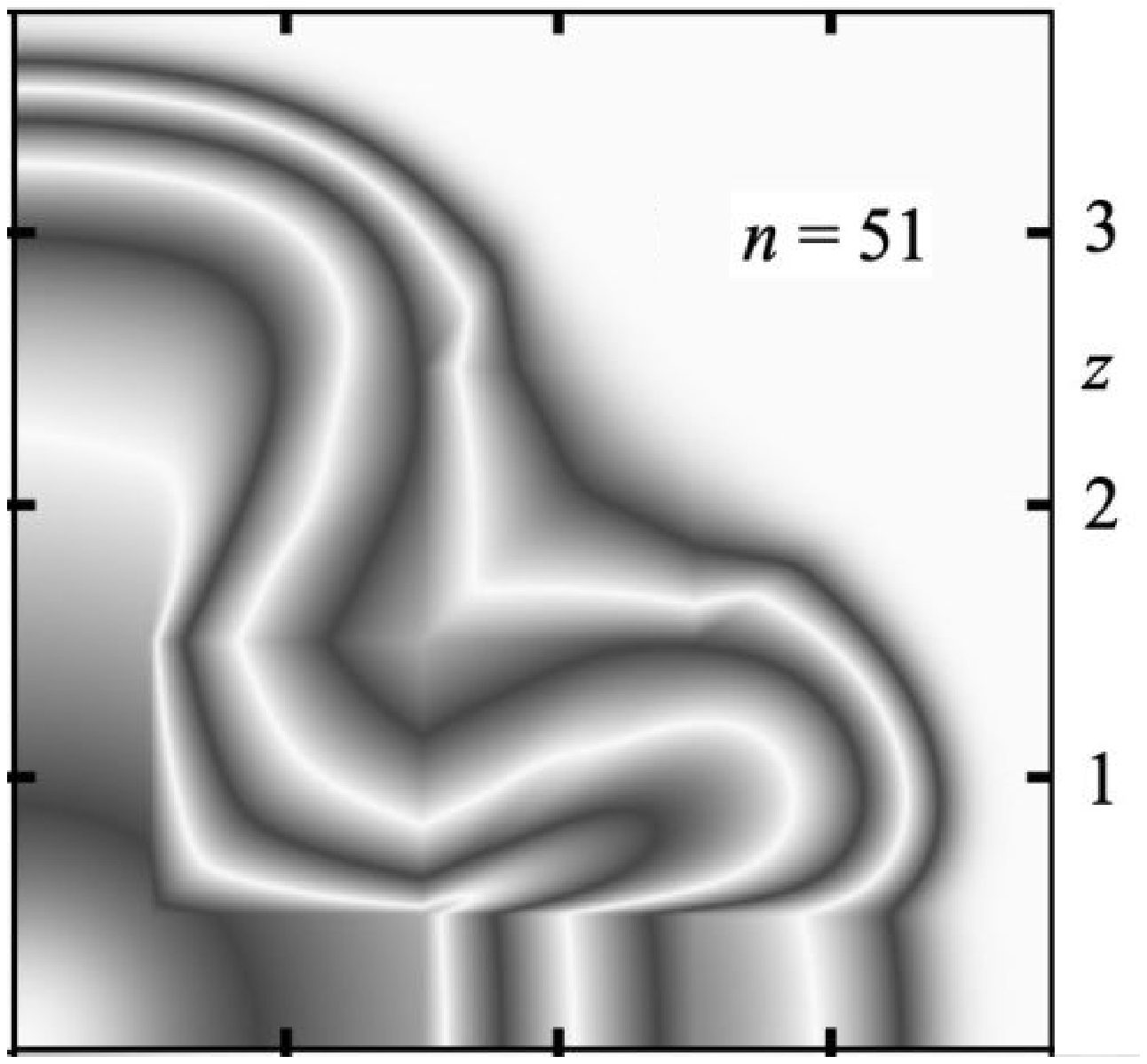}
(b)\includegraphics*[width=6.3cm]{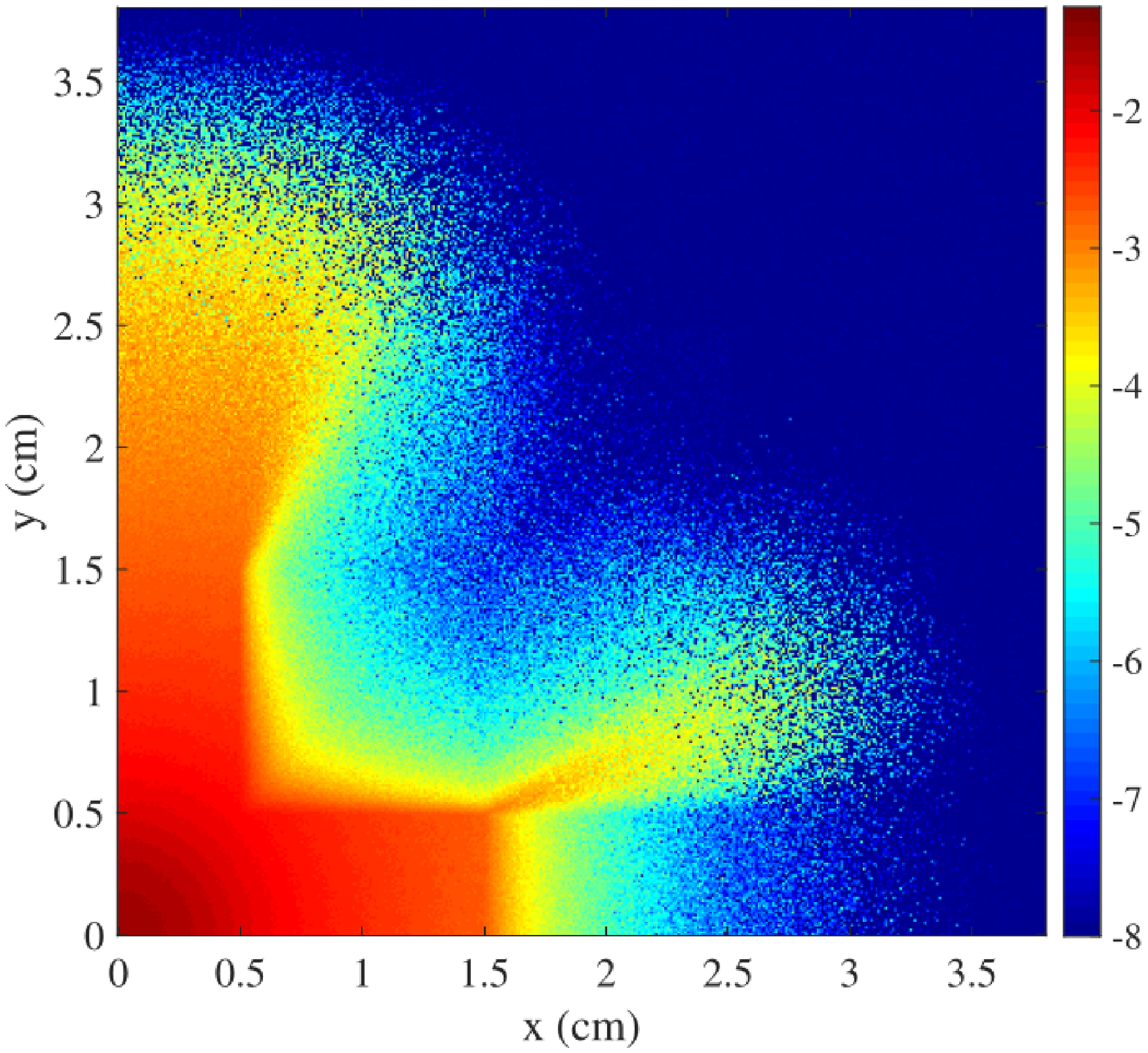}
(c)\includegraphics*[width=6.5cm]{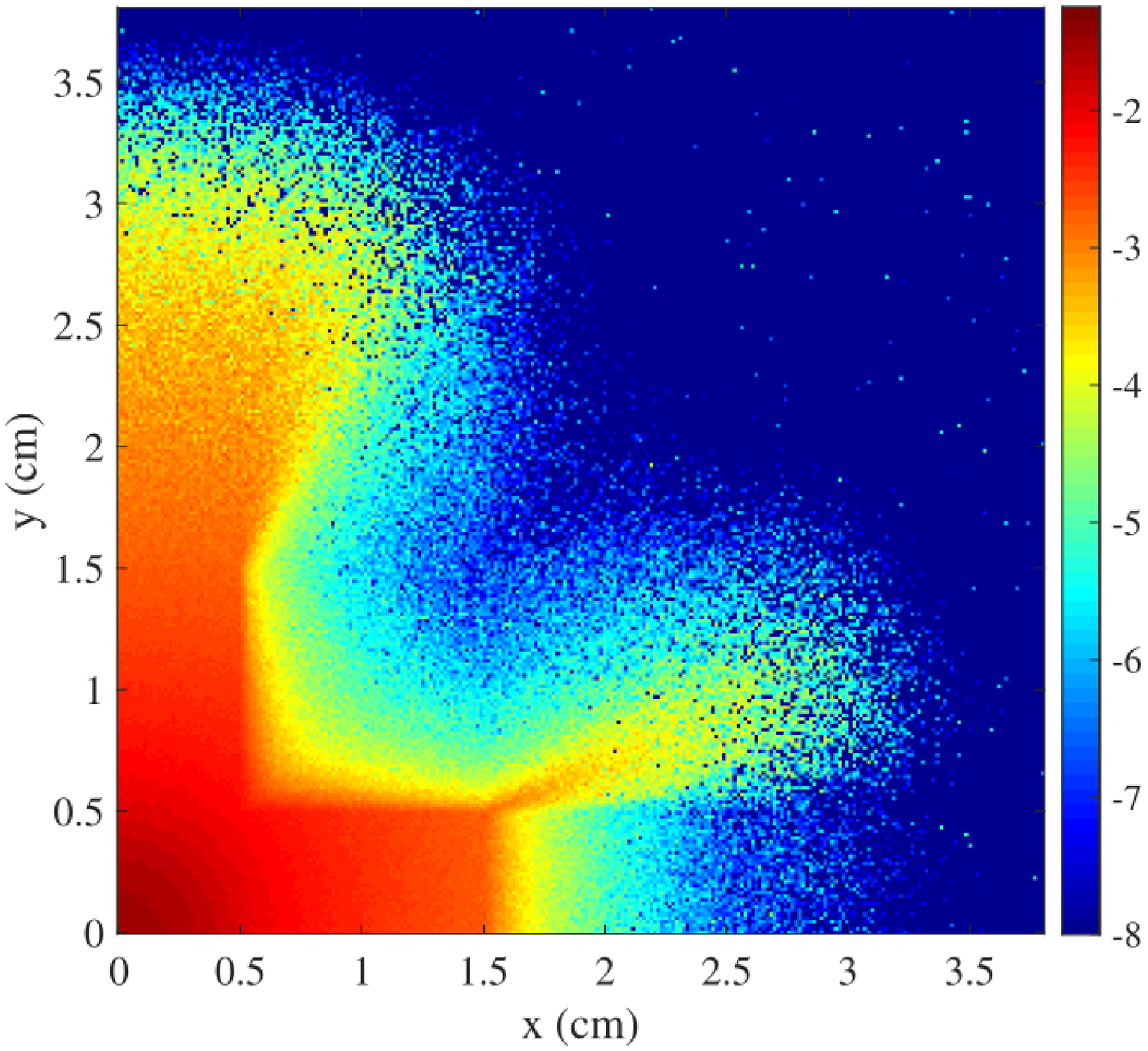}
(d)\includegraphics*[width=6.3cm]{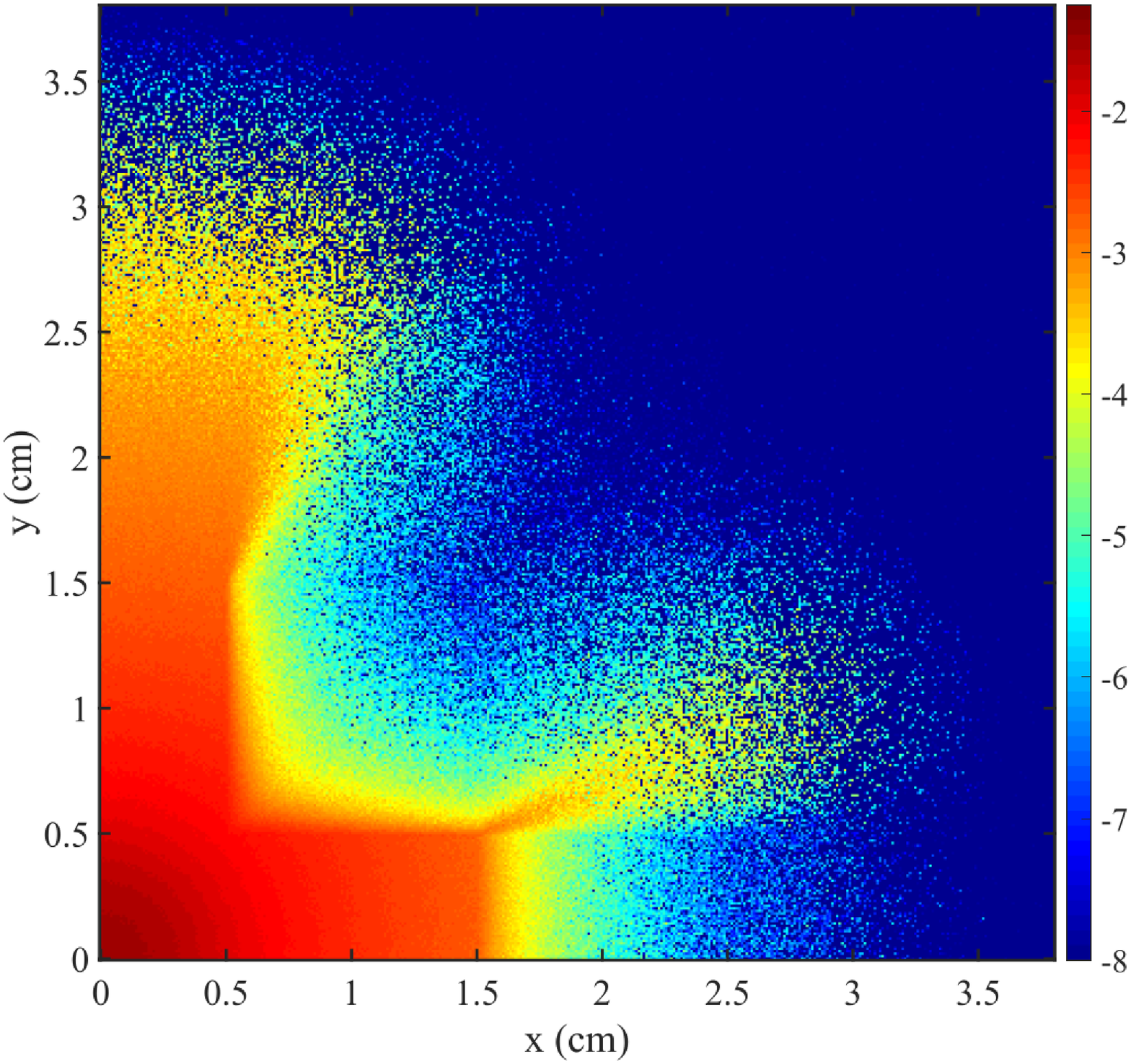}
\caption{(a) Contours of the logarithm of the radiation energy density at $ct=3$ using $P_{51}$ approximation. Fig (a) is taken from~\cite{olson2020}. (b) The radiation energy density at $ct=3$ using IMC method. (c) Same for ISMC method. (d) Same for DIMC method.}
\label{fig:olson20202D_rad}
\end{figure}
First, in Fig.~\ref{fig:olson20202D_rad}(b-d) we show a colormap of the radiation energy density in the entire domain as well as a reference from~\cite{olson2020} (Fig.~\ref{fig:olson20202D_rad}(a)). All MC methods, IMC, ISMC and DIMC agree very well with each other through the computational domain, and a good qualitative agreement with the $P_N$ reference data, including all geometrical patterns.

\begin{figure} 
(a)\includegraphics*[width=7.1cm]{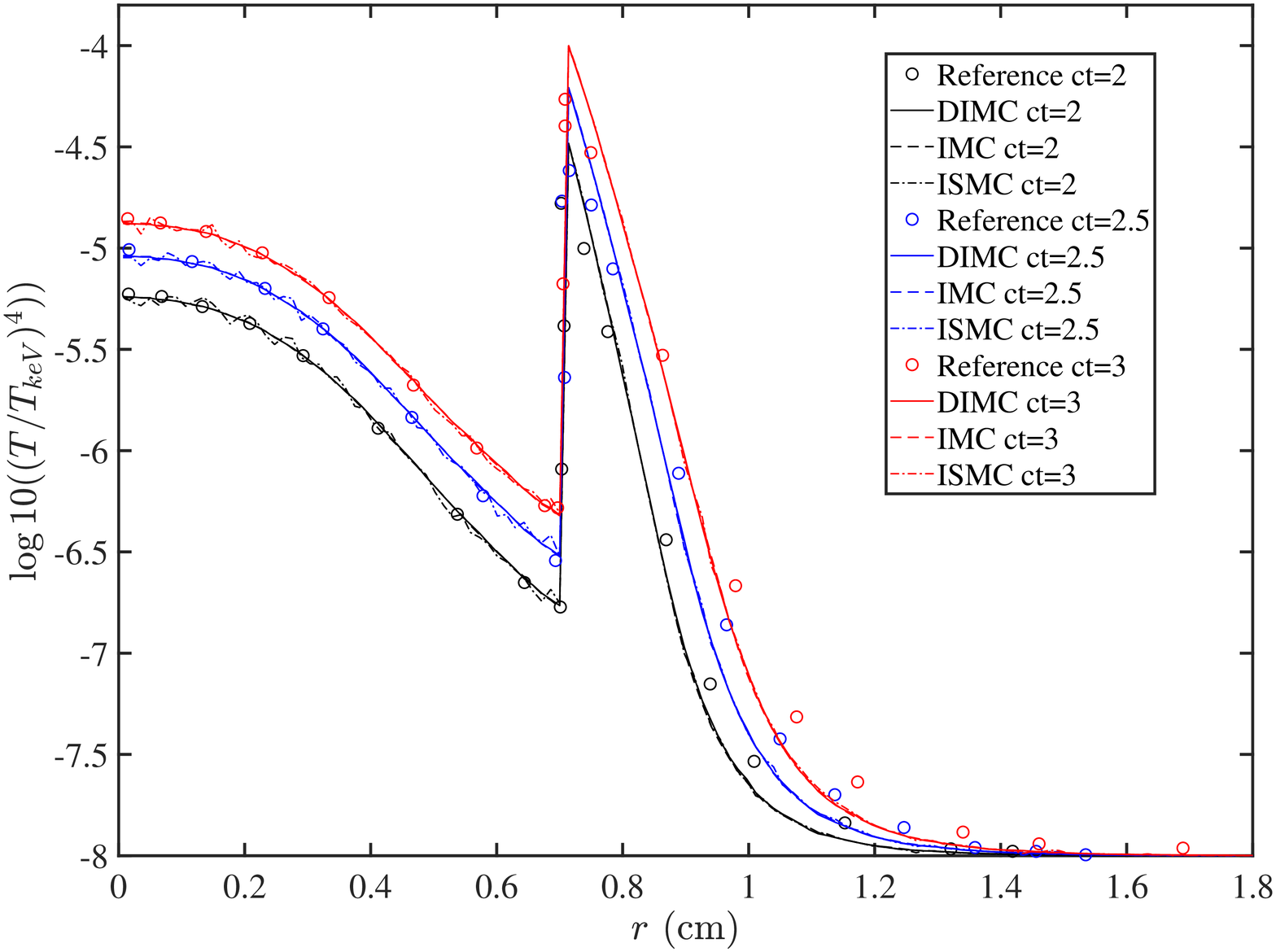}
(b)\includegraphics*[width=6.8cm]{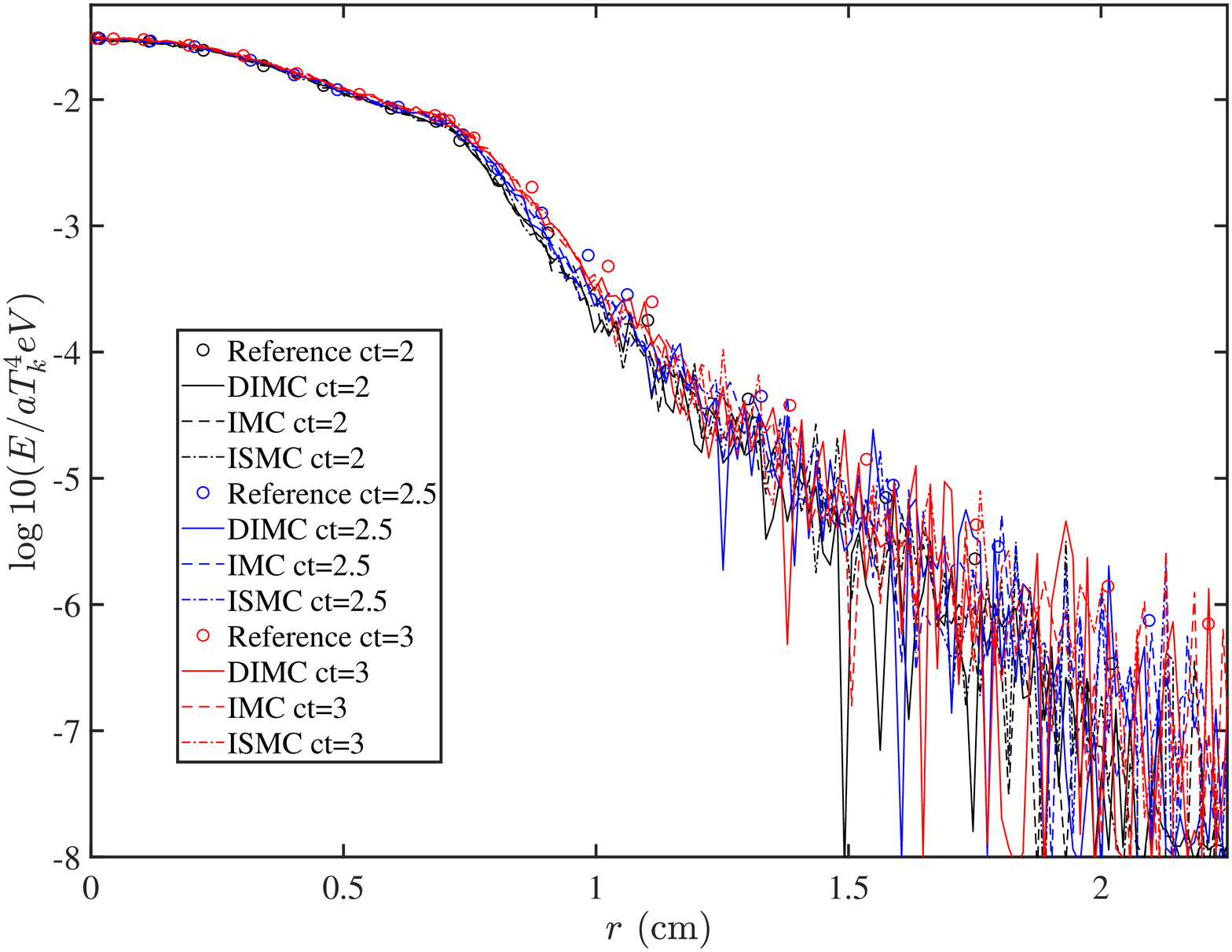}
\caption{(a) The material temperature at different times for the Olson~\cite{olson2020} 2D test problem. (b) The radiation energy density at different times for the Olson~\cite{olson2020} 2D test problem.}
\label{fig:olson20202D}
\end{figure}

Next, Fig.~\ref{fig:olson20202D} shows slices along the diagonal for (a) the material temperature and (b) the radiation energy density, along with reference $P_N$ data from~\cite{olson2020}. Again, there is a good agreement between all methods in the range $0\leqslant r\leqslant \sqrt{0.5}$. Farther away from the origin, the sharp discontinuity in the material properties causes a small difference between the MC methods and the reference solution. 

\section{Conclusions}
\label{discussion}

In this work we have presented a frequency-dependent (energy-dependent) generalization to the discrete implicit Monte-Carlo (DIMC) scheme that was introduced recently~\cite{Steinberg2}. The DIMC is a classic IMC implementation on one hand, but on the other hand uses the concept of two-kind of particles for avoiding teleportation errors, as introduced by Ahrens and Larsen~\cite{SMC} and later by Poëtte and Valentin~\cite{ISMC}. The main idea is when a photon continuously loses its energy due to absorption, a material particle is created, which is static. The new photons that emerge due to black-body emissions are sampled only from the locations of material particles, and thus, teleportation errors are avoided. DIMC has been checked before in gray opacity scenarios~\cite{Steinberg2}, and this work completes the generalization and the testing to frequency-dependent opacity problems. We have seen in the optically-thin problems, both in 1D and the 2D problems that the DIMC yields good results compared to the reference solution, and yields smoother results than ISMC, with comparable noise level as IMC. On the other hand, ISMC required two order of magnitudes more particles in order to converge at the same noise noise level. In optically-thick benchmarks, DIMC yields the best results, having no teleportation error even with a very low spatial resolution, like the ISMC results, while IMC needed a spatial resolution four times higher in order to achieve similar results.The results of the numerical examples, show no reason to think that the DIMC algorithms are biased, however, a consistent proof of unbiasedness is reserved for future work.
In addition, the benchmarks were run with {\em{continuous}} frequency dependency, but the results should be valid also for a finite number of groups, i.e., in multigroup problems, which could be another direction for future work.

\appendix
\section{Appendix: The fundamental DIMC algorithms}

In this appendix we present the fundamental algorithms of DIMC that were presented previously in the gray DIMC paper~\cite{Steinberg2}. An outline of the photon creation process is given in algorithm \ref{alg:create}. The full photon propagation algorithm is shown in algorithm \ref{alg:propogate}. The merging algorithm for the material particles is presented in algorithm \ref{alg:merge}.

\begin{algorithm}[t]
\SetAlgoLined
 \ForAll{\emph{cells}}
 {
  Nphoton = \emph{Number of photons to create in cell}\;
  V = Volume of cell\;
  Ephoton = $f\sigma_a c B(T)\Delta t$ V / Nphoton\;
  MaterialParticlesInCell = \emph{An array of all the material particles in the cell}\;
  \For{$i=1$ \KwTo \emph{Nphoton}}
  {
    photon = new Photon()\;
    idx = \emph{Draw a random index from \emph{MaterialParticlesInCell} weighted by \emph{MaterialParticlesInCell.energy}}\;
    photon.location = MaterialParticlesInCell[idx].location\;
    photon.direction = Sample a new direction\;
    photon.frequency = Sample a new frequency\;
    photon.energy = Ephoton\;
    MaterialParticlesInCell[idx].energy $-=$ Ephoton\;
    \emph{Subtract} Ephoton \emph{from cell's energy}\;
    // Rest of photon properties (i.e. angle, time) are determined as in IMC\ 
  }
 }
 \caption{Photon Creation}
\label{alg:create}
\end{algorithm}

\begin{algorithm}[t]
\SetAlgoLined
x = photon.location\;
EnergyForNewMaterialParticle = 0\;
 \While{\emph{photon}.\emph{time} $<$ \emph{time}  $+\Delta t$}
 {
 $r$ =  Random[0, 1]\;
 $d_{\mathrm{collision}}$ = $-\frac{\log{(1-r)}}{(1-f)\sigma_a(\nu)+\sigma_s(\nu)}$\;
 $d_{\mathrm{mesh}}$ = \emph{Distance for photon to reach cell boundary}\;
 $d_{\mathrm{time}}$ = \emph{speed of light} $\cdot$ (time + $\Delta t$ - photon.time )\;
 $d =$ min($d_{\mathrm{collision}}$, $d_{\mathrm{mesh}}$, $d_{\mathrm{time}}$)\;
    $\Delta E$ = photon.energy$\left(1-\exp{\left(-f\sigma_a(\nu) d\right)}\right)$\;
    $r$ =  Random[0, 1]\;
    ExpRandom = $-\log{\left(1+r\left(\exp{\left(-f\sigma_a(\nu) d\right)}-1\right)\right)} / f\sigma_a(\nu) d$\;
    $x$\textsubscript{create} = photon.location + ExpRandom$\cdot\mathbf{\Omega}\cdot d$\;
    photon.location $+= \mathbf{\Omega}\cdot d$\;
    $x =$ $\left(\text{EnergyForNewMaterialParticle}\cdot x + \Delta E \cdot x\text{\textsubscript{create}}\right)$ / $\left(\text{EnergyForNewMaterialParticle} + \Delta E\right)$\;
   EnergyForNewMaterialParticle += $\Delta E$\;
   photon.energy $-=$ $\Delta E$\;
 \eIf{$d_{\mathrm{collision}}=d$}
 {
    // Do as in IMC (i.e. update photon time, draw new angle, update cell energy)
  }
  {
   Create new material particle with energy EnergyForNewMaterialParticle and location x\;
        EnergyForNewMaterialParticle = 0\;
    // Rest as in IMC (i.e. update photon time, update cell energy)
 }
 }
 \caption{Single Photon Propagation}
 \label{alg:propogate}
 \end{algorithm}

\begin{algorithm}[t]
\SetAlgoLined
 \ForAll{cells}
 {
 MaterialParticlesInCell = \emph{An array of all the material particles in the cell}\;
  ParticlesToKeep = \emph{An array of length} N\textsubscript{keep} \emph{of material particles randomly selected weighted by their energy from} MaterialParticlesInCell\;
  \ForAll{\emph{m} = Material particle in \emph{MaterialParticlesInCell} but not in \emph{ParticlesToKeep}}
  {
    idx = \emph{Index of nearest particle in} ParticlesToKeep\;
    ParticlesToKeep[idx].location = $\left(\text{ParticlesToKeep}[\text{idx}].\text{location}\cdot \text{ParticlesToKeep}[\text{idx}].\text{energy} + \right.$ \
    $\left. \text{m}.\text{location}\cdot \text{m}.\text{energy}\right)$ / $\left(\text{ParticlesToKeep}[\text{idx}].\text{energy} + \text{m}.\text{energy}\right)$\;
    ParticlesToKeep[idx].energy $+=$ m.energy\;
   delete m\;
  }
 }
 \caption{Material Particle Merging}
\label{alg:merge}
\end{algorithm}

\bibliography{bibliography.bib}
\bibliographystyle{ieeetr}
\end{document}